\documentclass[aps,pra,twocolumn,superscriptaddress]{revtex4}

\usepackage{fancyhdr}     
\usepackage{amsfonts}
\usepackage{amsmath}
\usepackage[latin1]{inputenc}
\usepackage{graphicx,graphics,amssymb,times}
\usepackage{color}
\usepackage{natbib}
\usepackage{comment}

\newcommand{\rh}{\boldsymbol{ \rho}}

\begin{document}


\title{Optimising the use of detector arrays for measuring intensity correlations of photon pairs}

\begin{abstract}
Intensity correlation measurements form the basis of many experiments based on spontaneous parametric down-conversion. In the most common situation, two single-photon avalanche diodes and coincidence electronics are used in the detection of the photon pairs, and the coincidence count distributions are measured by making use of some scanning procedure. Here we analyse the measurement of intensity correlations using multi-element detector arrays. By considering the detector parameters such as the detection and noise probabilities, we found that the mean number of detected photons that maximises the visibility of the two photon correlations is approximately equal to the mean number of noise events in the detector array. We provide expressions predicting the strength of the measured intensity correlations as a function of the detector parameters and on the mean number of detected photons. We experimentally test our predictions by measuring far-field intensity correlations of spontaneous parametric down-conversion with an electron multiplying CCD camera, finding excellent agreement with the theoretical analysis. 
\end{abstract}

\author{Daniel S. Tasca}
\email[]{daniel.tasca@glasgow.ac.uk}
\address{SUPA, School of Physics and Astronomy, University of Glasgow, Glasgow, G12 8QQ, UK}

\author{Matthew P. Edgar}
\address{SUPA, School of Physics and Astronomy, University of Glasgow, Glasgow, G12 8QQ, UK}
 
\author{Frauke Izdebski}
\address{SUPA, School of Engineering and Physical Sciences, Heriot-Watt University, Edinburgh, EH14 4AS, UK }

\author{Gerald S. Buller}
\address{SUPA, School of Engineering and Physical Sciences, Heriot-Watt University, Edinburgh, EH14 4AS, UK }

\author{Miles J. Padgett}
\address{SUPA, School of Physics and Astronomy, University of Glasgow, Glasgow, G12 8QQ, UK}

\maketitle

\date{\today}

\section{Introduction}
Spontaneous parametric down-conversion (SPDC) is the most common technique used to produce entangled photon pairs. Due to the strong temporal correlation in the emission process, the photon pairs can be detected in coincidence between two different detectors using a very narrow time window \cite{Burnham70,Hong87}. The coincidence detection is an intensity correlation measurement: a coincidence event indicates the presence of one photon in each detector. Because most of the photo-detectors can not discriminate between one or more photons, measurements of intensity correlations (or single-photon counting) are realised with a low photon flux.

In general, the photon pairs from SPDC are correlated in many degrees of freedom (DOF). The process of observing these correlations depends on the DOF under consideration, but it is generally done by means of a scanning process. In the case of transverse linear position and momentum, correlations are measured by scanning the detectors across the detection planes, and coincidence counts are registered as a function of the detector positions. Due to the high dimensionality associated with the spatial DOF of the photons, a large number of scanning steps is normally necessary in order to measure the intensity correlations for a full set of spatial modes. Using scanning methods, spatial correlations have been measured for SPDC in many experiments \cite{Walborn10}. The scanning of the detectors across the detections planes has allowed the investigation of Einstein-Podolsky-Rosen (EPR) type correlations \cite{Howell04,Tasca09,Walborn11}, continuous variable entanglement \cite{Tasca08,Gomes09} and quantum key distribution \cite{Almeida05} using the transverse linear position and momentum of the down-converted photons. 

To take advantage of the high-dimensionality of the spatial DOF of single-photons, it is desirable to perform a projection onto a complete set of modes without the need of any scanning process. In order to implement a multimode detection of optical properties of single-photons, efficient and low-noise detector arrays become demanding. In the last years, detector arrays have been  increasingly used in measurements of SPDC light. The most common detector arrays used in SPDC experiments are the single-photon sensitive charge-coupled devices (CCD) such as the intensified CCD (ICCD) and the electron multiplying CCD (EMCCD). In 2002, Oemrawsingh {\it et al.} used an ICCD to measure far-field intensity correlations of SPDC light \cite{Oemrawsingh02}. In this experiment the authors post-selected images containing only two detected ``photons", thereby excluding most of the recorded data, and measured full-field intensity correlations of the photon pairs produced by SPDC.  Furthermore, there is a range of other experiments utilising cameras to detect the SPDC light, including measurements of transverse coherence properties \cite{Hamar10}, photon-number distributions \cite{perina12}, and sub-shot-noise correlations of intensity fluctuations \cite{Jedrkiewicz04,Blanchet08}.

Transverse spatial entanglement in SPDC is often detected via intensity correlation measurements in two conjugate planes of the SPDC crystal. Although most of these measurements have been realized by scanning single-photon detectors across the detection planes \cite{Howell04, DAngelo04,Tasca08, Tasca09, Gomes09}, it has recently been shown that EMCCDs are capable of measuring intensity correlations in both the near- and far-field of the SPDC source \cite{Edgar12,Moreau12}. This simultaneous access to the full transverse field of the photon pairs is a promising technique, creating new possibilities for the investigation of two photon entanglement and to applications in quantum information protocols. Before the works reported in Refs. \cite{Edgar12,Moreau12}, only far-field intensity correlations had been measured with detector arrays \cite{Oemrawsingh02,Zhang09,Blanchet10,Devaux2012}. Nevertheless, as the noise level in these single-photon sensitive cameras is still much higher than in single-photon avalanche diodes (SPAD), coincidence measurements using these cameras suffer from a much higher background than in the traditional scanning methods. Alternative approaches for measuring intensity correlations based on a digital micro-mirror array \cite{Dixon12,Howland13} or a time multiplexed fiber array \cite{Warburton2011,Leach2012} used in conjunction with a single-element detector have also recently been implemented.

While there has been discussion on the performance of detector arrays for single-photon discrimination \cite{Lantz08,Zhang09,Jedrkiewicz12} and low-light imaging \cite{Lantz08,Buchin11}, a detailed description of their performance in the measurement of low photon number intensity correlations is still lacking. In this paper we analyse the measurement of intensity correlations of photon pairs with multi-element detector arrays. We consider general detector parameters, such as the detection and noise probabilities, and give expressions for the measured intensity correlations. Given a certain level of noise in the detector arrays, we find the optimum photon flux in order to maximise the visibility of the two photon intensity correlations. In low-light imaging applications, the optimisation of the light level for single-photon discrimination leads to a photo-detection rate that is normally much higher than the dark-count rate of the detectors \cite{Lantz08,Jedrkiewicz12}. Nevertheless, in intensity correlation measurements, increasing the light level introduces undesired cross-correlations that scale with the square of the mean number of detected photons. For sufficiently low level of noise (of the order of 1 noise event every 100 pixels), we find that the mean number of detected photons that maximises the visibility of the intensity correlations is of the same order as the number of the noise events. 

The paper is organised as follows. In section \ref{sec:IC2photons} we briefly describe the intensity correlation function of a two photon state. In section \ref{sec:CCDarrays} we discuss the use of CCD arrays in the single-photon counting regime, defining the operational parameters that will be used in the derivation of our results. Our main findings are discussed in section \ref{sec:IC}. We start by defining the different contributions to the measured correlation function, and analyse how these different terms scale with the mean number of detected modes. We then provide general expressions for the measured correlation function and for its visibility as a function of the detector parameters and of the mean number of detected modes. Using the derived expressions, we calculate the mean number of detected modes that maximises the visibility of the two-photon intensity correlations. We finish by illustrating our results with an example. In section \ref{sec:Experiment} we report on an experiment using an EMCCD camera to measure intensity correlations in the far-field of a SPDC source. We apply our results considering specific parameters of  the EMCCD and provide measurements with excellent agreement with our theoretical predictions. We conclude in section \ref{sec:Conclusions}.

\section{Intensity correlations of two-photon states}

\label{sec:IC2photons}
Here we consider a two-photon state described by the density matrix $\hat{\varrho}_{12}$ and orthogonal  projective measurements onto a set of optical modes. These projective measurements of the two-photon state give rise to the coincidence counts distribution $C(\boldsymbol{\xi}_1,\boldsymbol{\xi}_2)$, where $\boldsymbol{\xi}_1$ ($\boldsymbol{\xi}_2$)  represent the modes in which the projections on photon 1 (2) are carried out. The coincidence counts distribution is proportional to the normally ordered second-order correlation function of the two-photon state \cite{Walborn10},
\begin{equation}\label{EQ:G2def}
C(\boldsymbol{\xi}_1,\boldsymbol{\xi}_2) \propto G^{(2)}(\boldsymbol{\xi}_1,\boldsymbol{\xi}_2)= \langle : \hat{N}(\boldsymbol{\xi}_1)\hat{N}(\boldsymbol{\xi}_2) : \rangle_{\hat{\varrho}_{12}},
\end{equation}
where $\hat{N}(\boldsymbol{\xi}_j) \equiv \hat{a}^{\dagger}(\boldsymbol{\xi}_j)\hat{a}(\boldsymbol{\xi}_j)$ is the number operator associated with the mode $\boldsymbol{\xi}_j$, with $\hat{a}(\boldsymbol{\xi}_j)$ and $\hat{a}^{\dagger}(\boldsymbol{\xi}_j)$ being respectively the annihilation and creation operators for the given mode. The average is taken over the two-photon quantum state $\hat{\varrho}_{12}$, and the symbol \, ``$: \, :$" \, indicates that the average is taken with the operators normally ordered. The second-order correlation function given in equation \eqref{EQ:G2def} gives the joint probability for the detection of the photon pairs in the modes $\boldsymbol{\xi}_1$ and $\boldsymbol{\xi}_2$.

Let us define the transverse coordinates $\rh_1=(x_1,y_1)$ and $\rh_2=(x_2,y_2)$ at the detection planes of photons $1$ and $2$, respectively. The spatial distribution of coincidence counts is proportional to the joint detection probability $\mathcal{P}(\rh_1,\rh_2)$, which is obtained by projecting the two-photon state in the eigenstates $\{|\rh\rangle\}$ of transverse linear position 
\begin{equation}\label{EQ:CClinear}
C(\rh_1,\rh_2) \propto G^{(2)}(\rh_1,\rh_2) =  \mathcal{P}(\rh_1,\rh_2).
\end{equation}

Let us consider the spatial degrees of freedom of the two-photon field from SPDC, assuming that the photons have a well defined polarization and are detected through narrow band-pass interference filters. Under these assumptions, and working in the thin crystal approximation \cite{Monken98}, the spatial structure of the post-selected two-photon field can be described by a pure state with a detection amplitude $\Psi(\rh_1,\rh_2)$ as
\begin{equation} \label{EQ:SPDC StateP}
|\Psi\rangle=\int \int d\rh_1 d\rh_2 \, \Psi(\rh_1,\rh_2) |\rh_1\rangle |\rh_2\rangle,
\end{equation}
where the states $|\rh_j\rangle$ represent non-normalizable states of a single-photon in the transverse position $\rh_j$. The two-photon detection amplitude $\Psi(\rh_1,\rh_2)$ is non-separable and, in general, the photon pairs from SPDC are highly spatially correlated \cite{Walborn10}. The joint detection probability for the two-photon state  \eqref{EQ:SPDC StateP}  at the detection planes reads $\mathcal{P}(\rh_1,\rh_2)=|\Psi(\rh_1,\rh_2)|^2$, whereas
the intensity of the down-converted light is proportional to the marginal detection probability distribution of the down-converted fields
\begin{equation}\label{EQ:MargSphoton}
I(\rh_i) \propto \mathcal{P}(\rh_i)=\int d \rh_j  \mathcal{P}(\rh_i,\rh_j).
\end{equation}
%

\section{Single-photon sensitive CCD arrays}
\label{sec:CCDarrays}

As mentioned in the introduction, single-photon sensitive CCD arrays such as ICCD and EMCCD cameras are the most common detector arrays used for single-photon detection. The typical active area of these CCD arrays is of the order of $1$cm$^2$, with a pixel size $s_p$ varying from around $10\mu$m to $20\mu$m. Since the fill-factor of these cameras is close to 100\%, meaning that the pixels are adjacent to each other, these cameras can provide up to one million single-photon sensitive individual pixels with quantum efficiencies of up to approximately $40\%$ for ICCDs and $90\%$ for EMCCDs.  

When working in the single-photon counting regime with such cameras, the output of each pixel must be thresholded in order to decide whether it corresponds to a photo-detection or not \cite{Lantz08}. After this binary thresholding, each pixel is assigned a value ``$0$" or ``$1$", where $1$ corresponds to either a photo-detection or a noise event. The calculations derived herein are expressed as a function of generic parameters that can be applied to any single-photon detector: noise probability $p_n$ and the detection probability $p_d$. 
The noise probability $p_n$ is defined as the probability that the thresholded detector output is $1$ in the absence of photons whereas $p_d$ gives the probability that the detector output is $1$ in the presence of a photon. Both these quantities are a function of the threshold used, and may depend on the operational parameters of the camera. 

We consider that the photon flux on the detector array is low, such that only a few pixels are illuminated in each frame. The pixels in the detector array are synchronized, such that a frame corresponds to the simultaneous shot of all of the detectors in the array.  The average number of detected photons per frame  depends on the photon flux and on the exposure time $\tau_e$. We define the average photon flux over the area $s_p^2$ of the $i${\footnotesize -th} pixel to be $\phi_{i}$. We assume that the mean number of photons $\mu_{i}$ is much less than unity, such that we can write $\mu_{i} = \phi_{i}\,\tau_e \ll 1$. In this limit of low mean number of photons, the average number of events (either photo-detections or noise) in that pixel can be written as
\begin{equation}\label{EQ:AveNumEvents}
\langle N_{i} \rangle =   p_d \, \mu_{i}  +  p_n(1-  p_d \, \mu_{i}).
\end{equation}
where $N_i$ is the thresholded pixel output ($0$ or $1$) for the $i${\footnotesize -th} pixel and the average $\langle N_i \rangle$ is taken over many frames of the detector array. The first term in equation  \eqref{EQ:AveNumEvents} represents the average number of detected photons whereas the second term is the average number of noise events. The probability of having a noise event is given by the product of the noise probability $p_n$ (in the absence of photons) and the probability of {\it not} having a photon. When the mean number of photons increases, such that the probability of having more than one photon on the pixel is significant, $\mu_{i}$ must be replaced with the probability to have at least one photon, as each pixel can not detect more than one photon per exposure time.

For sufficiently low mean number of photons $\mu_{i}$ and noise probability $p_n$, equation \eqref{EQ:AveNumEvents} can be approximated by the sum of the average number of detected photons and the noise probability: $\langle N_{i} \rangle \approx   p_d \, \mu_{i} + p_n$. For a mean number of photons and noise probability of the order of $\lesssim 10^{-2}$, the relative error between the average number of events given by equation \eqref{EQ:AveNumEvents} and this approximation is less than $1.5\%$. We note that the mean number of photons $\mu_{i}$ can be adjusted either by tuning the exposure time $\tau_e$ or by tuning the photon flux $\phi_{i}$. The noise probability $p_n$ is a characteristic of the detectors and, once the operational parameters (threshold, gain, etc) have been chosen, $p_n$ is fixed. For a typical EMCCD camera cooled to $-85^{\circ}$C, operating at maximum gain and using a suitable threshold \cite{Lantz08}, the noise probability is of the order of $10^{-2}$. 

\section{Intensity correlations with detector arrays}
\label{sec:IC}

\subsection{Photon and detection modes}

Detector arrays can be directly employed in the discrimination of linear momentum or position modes of single-photons \cite{Edgar12,Moreau12}. However, they can also be employed in the discrimination of other optical modes with the addition of a {\it mode sorting} element. Some examples of mode sorting devices are polarizing beam splitters (for polarisation modes), dispersive media (for frequency modes) and orbital angular momentum (OAM) mode sorters \cite{Berkhout10} (for spatial OAM modes). Once the optical modes of interest are mapped on to different linear momentum modes, detector arrays can be used for the discrimination of these modes. In this way, the detection of a photon by a given detector can be associated with the projection of the single-photon state on to a given optical mode $\boldsymbol{\xi}$. 

\begin{figure}[t]
\begin{center}
\includegraphics[width=3.5in]{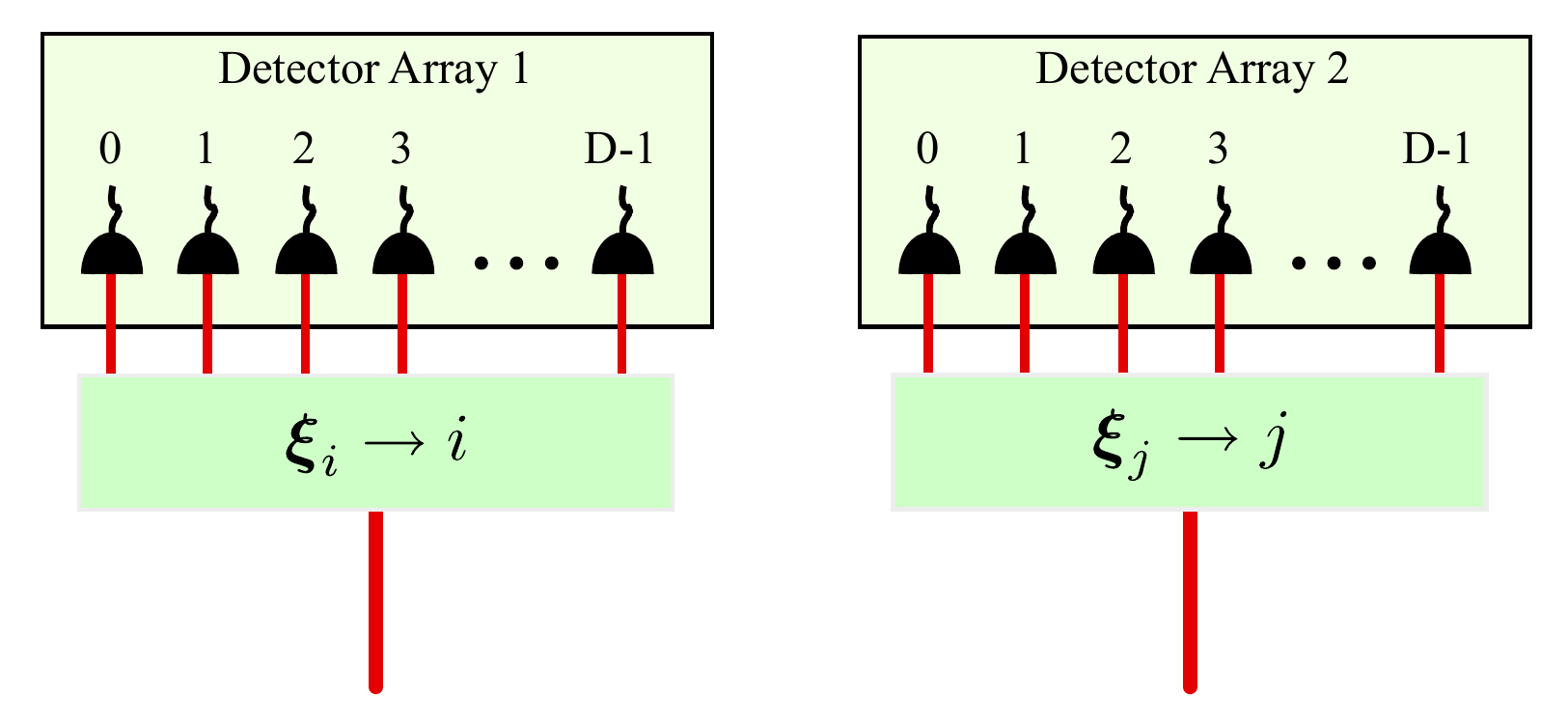} 
\caption{(Color online) Illustration of the detection of a photon pair with detector arrays. Each array contains $D$ detectors, whose elements are labeled $i$ (for photon $1$) and $j$ (for photon $2$). The two photon field generates a joint detection probability $\mathcal{P}_{ij}$ over the detector arrays, which we intend to measure by means of intensity correlations between the detectors of the two arrays.}
\label{DetectorArrays}
\end{center}
\end{figure}

Here we consider the measurement of intensity correlations of photon pairs in a configuration in which each photon field is detected by a different detector array, as illustrated in figure \ref{DetectorArrays}. The elements of the detector arrays of photon $1$ and $2$ are labeled $i$ and $j$, respectively.  The two-photon state $\hat{\varrho}_{12}$ generates a joint detection probability distribution $\mathcal{P}_{i j}$ over the detector arrays, which is associated with the second order correlation function $G^{(2)}(\boldsymbol{\xi}_1,\boldsymbol{\xi}_2)$ of the two-photon state (equation \ref{EQ:G2def}). Let us define the sorting operator  $\hat{\Pi} \equiv \sum_{{k},l} c_{{k}l} |{k} \rangle \langle \boldsymbol{\xi}_l |$ mapping a given optical mode $\boldsymbol{\xi}_l$ on to the $k${\footnotesize -th} detector with probability $|c_{{k}l}|^2$, and $\sum_k|c_{kl}|^2=1$.  For a {\it one-to-one} mapping, in which each mode is mapped on to a different detector, we have $c_{kl}=\delta_{kl}$. This one-to-one mapping gives the best discrimination between a photo-detection (associated with a given optical mode) and noise events without compromising the measurement resolution of the optical modes

Applying this mapping to the two photon state represented by the density matrix $\hat{\varrho}_{12}$, the joint detection probability of the  photon pair on the $i${\footnotesize -th} and $j${\footnotesize -th} elements of the detector arrays is given by 
\begin{equation}\label{EQ:Pijmapping}
\mathcal{P}_{ij}=\langle i| \langle j| \hat{\Pi}_1 \hat{\Pi}_2\, \hat{\varrho}_{12} \, \hat{\Pi}_1^{\dagger}\hat{\Pi}_2^{\dagger} \, | j \rangle | i \rangle.
\end{equation}
where the pair of optical modes $(\boldsymbol{\xi}_i,\boldsymbol{\xi}_j)$ is associated with the detection modes $(i,j)$.  For example, by using CCD arrays in the measurement of the continuous joint probability distribution of the photon pairs given in equation \eqref{EQ:CClinear}, one obtains a coarse grained (or pixelated) probability distribution $\mathcal{P}_{ij}$ of the two photon field over the CCD arrays, and the mapping is associated with the magnification of the optical system. Hereafter we will refer to the joint detection probability $\mathcal{P}_{ij}$ of the photon pair on the detector arrays, where $i$ and $j$ label the detector array elements or the detection modes associated with each detector.

\subsection{Coincidence count distributions}

The intensity correlation function is reconstructed from a series of frames by counting the coincidences as a function of the pixel coordinates (or labels $i$, $j$). We assume that the two detector arrays are synchronised as to detect photons belonging to the same generated pair. Each synchronised frame of the detector arrays has the duration of an exposure time $\tau_e$ and produces a list with the output of all detectors in the detector arrays. For each frame, the thresholded detector outputs (either $0$ or $1$) from one detector array are correlated with those from the other detector array, generating a coincidence count between the detectors $i$ and $j$ whenever the product $N_{i} \, N_{j}$ of the outputs is $1$. After acquiring a significant number of frames $N_F$, the coincidence counts distribution reads
\begin{equation}\label{EQ:CCsum}
C_{i j}=\sum_{k=1}^{N_F}C^k_{i j},
\end{equation}
where $k$ labels the frame. The measured intensity correlation function is obtained from the normalised coincidence counts distribution as
\begin{equation}\label{EQ:G2CC}
G^{(2)}_{i j}= \langle N_{i} \, N_{j} \rangle = \frac{C_{i j}}{\sum_{i j}C_{i j}},
\end{equation}

In each detector array, the number of events in a given frame is the sum of the number of detected photons and the number of noise events, according to equation \eqref{EQ:AveNumEvents}. In the calculation of the intensity correlations, each event is correlated with every other event. Apart from the {\it true} two-photon intensity correlation that is of interest, we also measure coincidences between noise and photo-detections and between noise and noise. Furthermore, if the number of detected pairs per frame is greater than one, there are also coincidence counts between photons of different pairs. All these unwanted coincidence counts contribute to decrease the visibility of the two-photon correlation. Summing the coincidence counts due to all frames according to equation \eqref{EQ:CCsum}, we obtain a coincidence count distribution  $C_{i j}$ which has four indistinguishable different contributions
\begin{equation}\label{EQ:CC4terms}
C_{i j} = (C_{pair})_{i j} + (C_{cross})_{i j} + (C_{n-n})_{i j} + (C_{n-ph})_{i j},
\end{equation} 
where each term has a different spatial distribution. The first term is the two-photon coincidence counts distribution of interest while the second term accounts for the cross coincidence counts between photons of different pairs. The number of cross coincidence counts is a function of the average number of photons detected per frame, or the average number of populated modes during an acquisition time. The third term accounts for the coincidences between noise events, and the fourth between photons and noise. 
The result of the measurement of the intensity correlation function \eqref{EQ:G2CC} is a {\it background lifted} version of the two-photon joint probability distribution $\mathcal{P}_{ij}$. For a pair of detectors $(i,j)$ for which the joint probability distribution of the photon pair is null, the measured coincidence count distribution is the sum of the three others terms in equation \eqref{EQ:CC4terms}. 

Let us define the visibility $\mathcal{V}$ of the measured intensity correlation \eqref{EQ:G2CC} as
\begin{equation}\label{EQ:Vdef}
\mathcal{V} \equiv  \frac{G^{(2)}_{i j} - \bar{G}^{(2)}_{i j}}{G^{(2)}_{i j} + \bar{G}^{(2)}_{i j}},
\end{equation}
where $\bar{G}^{(2)}_{i j}$ is defined as the intensity correlation \eqref{EQ:CC4terms} with the contribution from a photon pair, $(C_{pair})_{i j}$, equal to zero. We aim to maximise the visibility of the two-photon intensity correlation \eqref{EQ:Vdef}. In the ideal case in which the three unwanted coincidence count terms approach zero, the visibility of the two-photon correlation would approach unity. For real detectors with noise probability $p_n$, the visibility \eqref{EQ:Vdef} is decreased due to the coincidence counts introduced by noise events. In this situation, it is desirable to increase the average number of populated modes per frame, so that the contribution of the coincidence counts between pairs $(C_{pair})_{i j}$ to the total number of coincidences  \eqref{EQ:CC4terms} becomes statistically significant over the coincidence counts introduced by noise. On the other hand, increasing the number of populated modes also introduces unwanted correlations between photons and noise and between photons of different modes. Each of the four terms in equation \eqref{EQ:CC4terms} scales differently with the average number of populated modes. 

In practice, one is able to control the average number of events per frame by adjusting either the exposure time or the down-conversion source intensity. Although it is not possible to distinguish if a particular event is due to a photo-detection or noise, one is able to predict the average number of detected photons per frame in the full detector array. Because the noise probability $p_n$ is known, either from factory specification or from previous characterization, it is possible to predict how many of the events would correspond to photo-detection and how many to noise. What would then be the opitmal average number of detected photons in order to maximise the visibility \eqref{EQ:Vdef} of the two-photon intensity correlation? In order to answer this question, we consider how each term in equation \eqref{EQ:CC4terms} scales with the number of detected photons.

\subsection{Occupation probabilities of the detection modes}

For a continuous-wave (CW) SPDC source, the average number of down-converted pairs is proportional to time ($\tau_e$). In the case of a pulsed SPDC source, the average number of down-converted pairs is proportional to the number of pulses of the pump laser during one exposure time. In the limit where $\tau_e \gg R_r^{-1}$, where $R_r$ is the repetition rate of the pump laser, we also expect the average number of down-converted pairs to be proportional to $\tau_e$.  As the number of pump photons interacting within the non-linear crystal during a time $\tau_e$ is very large, and the down-conversion probability is very small, we can consider a Poissonian distribution $P_n=(e^{-\mu} \mu^{n}/n!)$ for the number of generated {\it pairs}, being $\mu=\langle n \rangle$ the mean number of emitted pairs.  

Considering $n$ down-converted pairs being emitted during one exposure time $\tau_e$, the joint probability distribution for the $n$ photon pairs over the detector arrays reads 
\begin{equation}\label{EQ:P_Npairs}
\mathcal{P}_{i'j',\dots,i^{\prime \dots \prime} j^{\prime \dots \prime}}  = \mathcal{P}^{(1)}_{i'j'} \times \mathcal{P}^{(2)}_{i''j''} \times \dots \times \mathcal{P}^{(n)}_{i^{\prime \dots \prime} j^{\prime \dots \prime}}.
\end{equation}
This probability represents the distribution of modes for $n$ independent photon pairs. Note that we are neglecting the probability of a four-photon emission, a process that increases the probability of having two pairs occupying the same mode due to stimulated emission \cite{Torren12}. In our case, we consider $n$ independent two-photon emissions during one exposure time of the detector arrays; for example, a pulsed pump laser with repetition rate of 100MHz generating an average of $10^{-3}$ down-converted pairs per pulse will give approximately 10 pairs during an exposure time of $\tau_e=0.1$ms. 

Each of the four terms in equation \eqref{EQ:CC4terms} is associated with a different occupation probability of the detection modes, which we calculate using the $n$ photon-pairs probability distribution given by equation \eqref{EQ:P_Npairs}.

\subsubsection{Pair occupation probability}

We begin by calculating the probability that  {\it at least} one out of $n$ generated pairs is emitted in the mode corresponding to the pair of detectors $(i, j)$. We define this conditional probability as $\mu^{pair}_{i j | n}$. Using equation \eqref{EQ:P_Npairs} to sum over all these possibilities we get
\begin{equation}\label{EQ:ModeNumber}
\mu^{pair}_{i j | n} = \mathcal{P}_{i j}  \times \sum_{k=0}^{n-1} (1- \mathcal{P}_{i j} )^k= 1-(1- \mathcal{P}_{ij})^n. 
\end{equation}
This conditional probability gives the probability of occupancy of the mode $(i ,j)$ by a pair of photons given that $n$ pairs were emitted during one exposure time $\tau_e$.  The mean probability of occupancy of detector modes $(i ,j)$ by a pair is obtained by averaging $\mu^{pair}_{i j | n}$ with the probability distribution $P_n$ for the number of generated pairs,
\begin{equation}\label{EQ:ModeNumberAve}
\mu^{pair}_{i j} \equiv \sum_{n=0}^{\infty} P_n \, \mu^{pair}_{i j | n},
\end{equation}
where we use the Poissonian distribution $P_n$ for the number of down-converted pairs. The quantity $\mu^{pair}_{i j}$ is the probability that the detection mode $(i,j)$ is populated with at least one photon pair. From equation \eqref{EQ:ModeNumberAve} we can calculate the average number of {\it populated} modes $\mu_p$ by summing over all modes; $\mu_p \equiv \sum_{i,j}  \mu^{pair}_{i j}$. We note that the mean number of emitted pairs $\mu$ is greater than the mean number of populated modes ($\mu_p< \mu$), as the source can emit more than one pair in the same mode.

\subsubsection{Cross occupation probability}

The cross coincidence count rate depends on the probability that a photon from a pair is occupying the detector mode $i$ and a photon from {\it another} pair is occupying the detector mode $j$. The marginal probability distribution $\mathcal{P}_{i}\equiv\sum_{j'} \mathcal{P}_{ij'}$ gives the probability that a photon from a given pair is occupying the detector mode $i$. Nevertheless, this marginal probability also includes the probability that its correlated photon is in the mode $j$, $\mathcal{P}_{i j}$, whose contribution has already been taken into account in $\mu^{pair}_{i j}$  \eqref{EQ:ModeNumberAve}. Then, the joint occupation of the detector modes $i$ and $j$ by photons of different pairs that are not in the modes $(i, j)$ depends on the product of the probabilities $\sum_{j' \neq j} \mathcal{P}_{ij'} =(\mathcal{P}_{i}-\mathcal{P}_{ij})$ and $\sum_{i' \neq i} \mathcal{P}_{i j} =(\mathcal{P}_{j}-\mathcal{P}_{ij})$. Using the $n$ photon pairs probability distribution \eqref{EQ:P_Npairs} to sum over all these possibilities, we calculate this probability to be
\begin{eqnarray}\label{EQ:mucrossij}
\mu^{cross}_{i j|n} &=&  ( \mathcal{P}_{i} - \mathcal{P}_{ij}) ( \mathcal{P}_{j} - \mathcal{P}_{ij}) \times \sum_{k=0}^{n-2} (1-\mathcal{P}_{ij})^k  \\
\nonumber &\times&     \sum_{l=0}^{n-2-k} \left[ (1-\mathcal{P}_{i})^l (1- \mathcal{P}_{i} - \mathcal{P}_{j}+ \mathcal{P}_{ij})^{n-2-k-l} \right. \\
\nonumber   && \phantom{\sum_{l=0}^{n-2-}}  +  \left. (1-\mathcal{P}_{j})^l  (1- \mathcal{P}_{i} - \mathcal{P}_{j}+ \mathcal{P}_{ij})^{n-2-k-l} \right]  \\
\nonumber &=&(1- \mathcal{P}_{ij})^n + (1- \mathcal{P}_{i} - \mathcal{P}_{j}+ \mathcal{P}_{ij})^n -(1- \mathcal{P}_{i})^n \\ 
\nonumber &-&  (1- \mathcal{P}_{j})^n,
\end{eqnarray}
where we have defined $\mu^{cross}_{i j|n}$ as the conditional probability of occupancy of modes $i$ and $j$ by photons of different pairs given that $n$ photons pairs were emitted. As before, we take the average of $\mu^{cross}_{i j|n}$ with the the probability $P_n$ for the emission of $n$ pairs to obtain the mean occupation number of detectors $i$ and $j$ by photons of different pairs
\begin{equation}\label{EQ:ModeNumberAveCross}
\mu^{cross}_{i j} \equiv \sum_{n=0}^{\infty} P_n \, \mu^{cross}_{i j | n}.
\end{equation}
%

\subsubsection{Photon-noise occupation probabilities}

The coincidence count rate between a noise event and a photo-detection is proportional to the probability that at least one photon is detected by detector $i$ and none by detector $j$, and {\it vice-versa}. Given that $n$ pairs were emitted during one exposure time $\tau_e$, we use equation \eqref{EQ:P_Npairs} to calculate the probability $\mu_{i \bar{j}|n}$ that at least one photon is occupying the detector mode $i$ but none is occupying the detector mode $j$ to be
\begin{eqnarray}\label{EQ:muYesBar}
 \nonumber \mu_{i \bar{j}|n} &=& (\mathcal{P}_{i} - \mathcal{P}_{ij})  \sum_{k=0}^{n-1}  (1-\mathcal{P}_{j})^k  (1-\mathcal{P}_{i}-\mathcal{P}_{j}+\mathcal{P}_{ij})^{n-1-k} \\
 &=& (1-\mathcal{P}_{j})^n-(1-\mathcal{P}_{i}-\mathcal{P}_{j}+\mathcal{P}_{ij})^n.
\end{eqnarray}
Analogously, the probability $\mu_{\bar{i} j|n}$ for at least one photon on detector $j$ and none on detector $i$ is
\begin{eqnarray}\label{EQ:muBarYes}
 \nonumber \mu_{\bar{i} j|n} &=& (\mathcal{P}_{j} - \mathcal{P}_{ij})  \sum_{k=0}^{n-1}  (1-\mathcal{P}_{i})^k  (1-\mathcal{P}_{i}-\mathcal{P}_{j}+\mathcal{P}_{ij})^{n-1-k} \\ 
 &=&(1-\mathcal{P}_{i})^n-(1-\mathcal{P}_{i}-\mathcal{P}_{j}+\mathcal{P}_{ij})^n,
\end{eqnarray}
and the means are obtained by taking the average with the probability $P_n$
\begin{equation}\label{EQ:ModeNumberPhotonNoise}
\mu_{i \bar{j}} \equiv \sum_{n=0}^{\infty} P_n \, \mu_{i \bar{ j}|n},
\end{equation}
and  
\begin{equation}\label{EQ:ModeNumberNoisePhoton}
\mu_{\bar{i} j} \equiv \sum_{n=0}^{\infty} P_n \, \mu_{\bar{i} j|n}.
\end{equation}
%

\subsubsection{Noise-noise occupation probability}

Finally, a coincidence count between two noise events requires that neither detector $i$ nor $j$ detect a photon. The probability $\mu_{\bar{i} \bar{j}|n}$ that no photon is occupying these detection modes given that $n$ pairs are emitted is
\begin{equation}\label{EQ:muBarBar}
\mu_{\bar{i} \bar{j}|n}=  (1-\mathcal{P}_{i}-\mathcal{P}_{j}+\mathcal{P}_{ij})^n,
\end{equation}
and the mean is given by
\begin{equation}\label{EQ:ModeNumberNoiseNoise}
\mu_{\bar{i} \bar{j}} \equiv \sum_{n=0}^{\infty} P_n \, \mu_{\bar{i} \bar{j}|n}.
\end{equation}

In the following, the occupation probabilities defined in equations \eqref{EQ:ModeNumberAve}, \eqref{EQ:ModeNumberAveCross}, \eqref{EQ:ModeNumberPhotonNoise} , \eqref{EQ:ModeNumberNoisePhoton} and \eqref{EQ:ModeNumberNoiseNoise} will  be used in the calculation of the coincidence count rates involved in the measurement of the intensity correlations of the photon pairs. It is interesting to notice that the conditional probabilities associated with these means sum to one, $\mu^{pair}_{i j|n} + \mu^{cross}_{i j|n} + \mu_{i \bar{j}|n} + \mu_{\bar{i} j|n} + \mu_{\bar{i} \bar{j}|n}=  1$, as it should be by their definition.

\subsection{Ideal case: unity detection probability and noiseless detectors}

Let us first consider the ideal case of unity detection probability ($p_d=1$) and no noise events ($p_n=0$). For $p_d=1$, and since each pair of detectors can only measure one coincidence count per frame, the coincidence count rates $(C_{pair})_{i j} $ and $(C_{cross})_{i j}$ are directly proportional to the occupation probabilities $\mu^{pair}_{i j}$ \eqref{EQ:ModeNumberAve} and  $\mu^{cross}_{i j}$ \eqref{EQ:ModeNumberAveCross}, respectively. The measured correlation function reads
\begin{equation}\label{EQ:G2p=1DEF}
G^{(2)}_{i j} \propto \mu^{pair}_{i j} + \mu^{cross}_{i j}.
\end{equation}

In this ideal limit, it is easy to see that the optimum photon flux to work with is such that the average number of detected pairs per frame is much less than one: $\mu \ll1$. In this case, many of the acquired frames will contain no photons, from which no coincidences will be registered. Frames containing one pair will generate one coincidence count with spatial distribution given by the joint detection probability $\mathcal{P}_{i j}$ of the photon pair on the detector arrays. Frames containing more than one pair will generate coincidence counts between photon pairs and also between photons of different pairs. As the probability that  a given mode is populated with more than one photon pair also decreases with the mean number of emitted pairs, we can assume that the mean number of populated modes $\mu_p$ will be approximately the same as the mean number of emitted pairs: $\mu_p \approx \mu$. More specifically, this approximation holds as long as the number of modes for which the photon pairs have non-zero joint detection probability is much greater than the mean number of emitted pairs. In this situation, we can approximate the conditional probabilities given in equations \eqref{EQ:ModeNumber} and \eqref{EQ:mucrossij} by $\mu^{pair}_{i j|n} \approx  n \mathcal{P}_{i j}$ and $\mu^{cross}_{i j|n} \approx  n(n-1) \,(\mathcal{P}_{i} - \mathcal{P}_{i j})(\mathcal{P}_{j} - \mathcal{P}_{i j})$, respectively.  Using these approximations, we can write the measured correlation function given in equation \eqref{EQ:G2p=1DEF} as
\begin{equation}\label{EQ:G2p=1}
G^{(2)}_{i j} \approx \mu \, \mathcal{P}_{i j} +  \mu^2 \,(\mathcal{P}_{i} - \mathcal{P}_{i j})(\mathcal{P}_{j} - \mathcal{P}_{i j}),
\end{equation}
where we have used $\langle n(n-1) \rangle=\langle n \rangle^2=\mu^2$ for the Poissonian distribution $P_n$. We can see that whilst the coincidence counts between pairs scales linearly with $\mu$, the cross coincidence counts scales with its square. Also, the contribution of the cross coincidence counts to the intensity correlation function is smaller on pairs of detectors for which the joint detection probability is larger. For a pair of detectors for which $\mathcal{P}_{i j}=0$, the only contribution to the measured correlation function \eqref{EQ:G2p=1} is due to cross coincidence counts, whose probability is given by the product of the individual detection probabilities in each detector.

Using the visibility defined in equation \eqref{EQ:Vdef}, we can write the visibility of the correlation function \eqref{EQ:G2p=1} as
\begin{equation}\label{EQ:Vp=1}
\mathcal{V} \approx \frac{ \mathcal{P}_{i j}[1 -  \mu\, ( \mathcal{P}_{i} + \mathcal{P}_{j} - \mathcal{P}_{i j}) ]}{\mathcal{P}_{i j}[1 -  \mu\, ( \mathcal{P}_{i} + \mathcal{P}_{j} - \mathcal{P}_{i j}) ]+ 2\mu \,\mathcal{P}_{i} \mathcal{P}_{j}},
\end{equation}
where we have used $\bar{G}^{(2)}_{i j}= \mu^2 \,\mathcal{P}_{i}\mathcal{P}_{j}$. 

\subsection{Effect of reduced detection probability}

The effect of non-perfect detection probability $p_d <1$ on the measured intensity correlation \eqref{EQ:G2p=1DEF} can be incorporated by multiplying the joint ($\mathcal{P}_{ij}$) and marginals ($\mathcal{P}_{i}$ and $\mathcal{P}_{j}$) detection probabilities of the photon pair by $p_d^2$ and $p_d$, respectively. The probability to detect a photon pair is given by $p_d^2$, while the probabilities of detecting only one  or none of the photons from a pair is $2p_d(1-p_d)$ and $(1-p_d)^2$, respectively. It is worth noticing that a new type of cross coincidence count is introduced when $p_d <1$, namely when more than one photon pair is emitted in the same mode but the detected photons belong to different pairs. This effect is also incorporated in the coincidence count rates when introducing the detection probability $p_d$ in equations \eqref{EQ:ModeNumber} and \eqref{EQ:mucrossij}. For the low mean photon pair number approximation, the average number of {\it detected} photons in each detector array is given by $p_d\mu$, and we can write the measured intensity correlation as
\begin{equation}\label{EQ:G2p2}
G^{(2)}_{i j} \approx \mu \, p^2_d \, \mathcal{P}_{i j}  +  \mu^2 \, p^2_d (\mathcal{P}_{i} - p_d \, \mathcal{P}_{i j}) (\mathcal{P}_{j} - p_d \, \mathcal{P}_{i j}).
\end{equation}
Introducing the reduced detection probability alone (without noise) does not affect  the scaling of the visibility with the mean number of photon pairs $\mu$. Nevertheless, a larger number of frames $N_F'=N_F/p_d^2$ will be necessary in order to have the same statistical sampling of the ideal case of $p_d=1$. In this case, the visibility reads
\begin{equation}\label{EQ:SBRp<1}
\mathcal{V} \approx \frac{ \mathcal{P}_{i j}[1 -  p_d \mu\, ( \mathcal{P}_{i} + \mathcal{P}_{j} - p_d\, \mathcal{P}_{i j}) ]}{\mathcal{P}_{i j}[1 -  p_d\mu\, ( \mathcal{P}_{i} + \mathcal{P}_{j} - p_d\,\mathcal{P}_{i j}) ]+ 2\mu \,\mathcal{P}_{i} \mathcal{P}_{j}}.
\end{equation}

\subsection{Effect of reduced detection and increased noise probabilities}

Noise events contribute to the measured intensity correlation function in two ways: they generate coincidences with other noise events and coincidences with photo-events. The probability for a noise event to happen in a particular detector is given by the product of the noise probability $p_n$ and the probability that no photo-detection has taken place in the detector. Yet the probability for a noise event in detector $i$ depends only if this particular detector has not detected a photon, a coincidence count between this noise event and another event in detector $j$ depends on whether there is a photo-detection or another noise event on detector $j$. As the photo-detection probabilities on these detectors are correlated through the joint probability distribution $\mathcal{P}_{ij}$, the lack of  photo-detections on two detectors  is also correlated. In other words, coincidence counts between photo-detections and noise events are more likely to happen between a pair of detector for which the joint detection probability is smaller. On the other hand, coincidence counts between two noise events are more likely to happen on a pair of detectors for which the joint detection probability is greater. These effects are expressed in the occupation probabilities given in equations \eqref{EQ:ModeNumberPhotonNoise}, \eqref{EQ:ModeNumberNoisePhoton} and \eqref{EQ:ModeNumberNoiseNoise}.

We assume that noise probability $p_n$ is flat over the detector arrays and that, in the absence of photons, the noise events in different detectors are uncorrelated from one another. This means that the noise contributions to the measured intensity correlation function are given by the product of the occupation probabilities \eqref{EQ:ModeNumberPhotonNoise} and \eqref{EQ:ModeNumberNoisePhoton} with $p_n$ and of the occupation probability \eqref{EQ:ModeNumberNoiseNoise} with $p_n^2$.
Incorporating these contributions to the intensity correlation function  \eqref{EQ:G2p=1DEF}, we have 
\begin{equation}\label{EQ:G2ALL}
G^{(2)}_{i j} \propto  \mu^{pair}_{i j} + \mu^{cross}_{i j} + (\mu_{\bar{i} j} + \mu_{i \bar{j}}) \, p_n + \mu_{\bar{i} \bar{j}} \,  p_n^2.
\end{equation}
The total intensity correlation function \eqref{EQ:G2ALL}  has now four terms, each of which representing the four different contributions to the coincidence counts. Making use of the the conditional probabilities expressed in equations \eqref{EQ:ModeNumber}, \eqref{EQ:mucrossij}, \eqref{EQ:muYesBar}, \eqref{EQ:muBarYes} and \eqref{EQ:muBarBar}, we can write the measured intensity correlation explicitly as
\begin{eqnarray}\label{EQ:G2ALL2}
G^{(2)}_{i j} &\propto& \sum_{n=0}^{\infty} P_n \\
\nonumber &\times& \left\{ 1-  (1-p_n) [ (1- p_d \, \mathcal{P}_{i})^n + (1- p_d \,\mathcal{P}_{j})^n]  \right.    \\
\nonumber  && + \left.   (1-p_n)^2  (1- p_d \, \mathcal{P}_{i}-p_d \, \mathcal{P}_{j} + p_d^2 \, \mathcal{P}_{ij})^n \right\}.
\end{eqnarray}

As we did for the case of $p_n=0$, we shall now analyse the measured intensity correlation function \eqref{EQ:G2ALL2} in the limit of low mean photon pair number $\mu$. In this limit, we can approximate the conditional probabilities given in equantions \eqref{EQ:muYesBar}, \eqref{EQ:muBarYes} and \eqref{EQ:muBarBar} by $\mu_{i \bar{j}|n} \approx n (\mathcal{P}_{i} - \mathcal{P}_{ij})$, $\mu_{\bar{i} j|n} \approx n (\mathcal{P}_{j} - \mathcal{P}_{ij})$ and  $\mu_{\bar{i} \bar{j}|n} \approx 1$, respectively.  Introducing these approximations into the intensity correlation function \eqref{EQ:G2ALL}, we obtain
\begin{eqnarray}\label{EQ:G2AllApprox}
G^{(2)}_{i j} &\approx& \mu \, p^2_d \, \mathcal{P}_{i j}  +  \mu^2 \, p^2_d \, (\mathcal{P}_{i} - p_d \, \mathcal{P}_{i j}) (\mathcal{P}_{j} - p_d \, \mathcal{P}_{i j}) \\
 \nonumber &+& \mu \,p_d \,p_n \, [ (\mathcal{P}_{i} - p_d \, \mathcal{P}_{i j}) + (\mathcal{P}_{j} - p_d \, \mathcal{P}_{i j}) ] +p_n^2.
\end{eqnarray}
Equation \eqref{EQ:G2AllApprox} describes the measured intensity correlation function for the two-photon state as a function of the detector arrays parameters $p_d$ and $p_n$ in the limit of a low mean number of photon pairs $\mu$. Each of the four terms involved in equation \eqref{EQ:G2AllApprox} is a quadratic approximation of the exact terms shown in equation \eqref{EQ:G2ALL2} which are valid as long as $\mu_p \approx \mu$. As the mean number of emitted photon pairs $\mu$ gets larger, these quadratic approximations overestimate the corresponding exact probabilities. 

Using equation \eqref{EQ:G2AllApprox} with $\mathcal{P}_{i j}=0$, we get the correlation function $\bar{G}^{(2)}_{i j}=  \mu^2 \, p_d^2 \, \mathcal{P}_{i}\mathcal{P}_{j} + \mu \, p_d\, p_n \, (\mathcal{P}_{i} + \mathcal{P}_{j}) + p_n^2$ that is obtained for uncorrelated photon pairs with the same marginal probabilities $\mathcal{P}_i$ and  $\mathcal{P}_j$. Using this $\bar{G}^{(2)}_{i j}$, we calculate the visibility \eqref{EQ:Vdef} of the intensity correlation function \eqref{EQ:G2AllApprox} to be
\begin{widetext}
\begin{equation}\label{EQ:Vmain}
\mathcal{V} \approx \frac{ \mu \, p^2_d \, \mathcal{P}_{i j} [1  - p_d \, \mu \, (  \mathcal{P}_{i} + \mathcal{P}_{j} -p_d \, \mathcal{P}_{i j})  - 2p_n ] }{  \mu \, p^2_d \, \mathcal{P}_{i j} [1  - p_d \, \mu \, (  \mathcal{P}_{i} + \mathcal{P}_{j} -p_d \, \mathcal{P}_{i j})  - 2p_n ] + 2 p_d^2 \, \mu^2 \,  \mathcal{P}_{i} \mathcal{P}_{j}  + 2\mu \,p_d \,p_n \,( \mathcal{P}_{i}  + \mathcal{P}_{j} ) + 2p_n^2}.
\end{equation}
\end{widetext}
The visibility of the two-photon intensity correlation \eqref{EQ:Vmain} is now written as a function of the photon pair flux $\mu$, on the detection probability $p_d$ and on the noise probability $p_n $. In order to know the mean number of pairs that maximises the visibility in \eqref{EQ:Vmain}, we take its derivative with respect to $\mu$ and equate to zero. This calculation leads to 
\begin{equation}\label{EQ:Derivative=0}
\frac{\partial}{\partial \mu} \mathcal{V} =0 \, \Rightarrow \, ( p_d \mu \, \mathcal{P}_{i} ) ( p_d \mu \, \mathcal{P}_{j} ) \approx p_n^2,
\end{equation}
where we have considered only the terms up to third order in the product of the probabilities $p_n$,  $p_d^2\, \mathcal{P}_{ij}$, $p_d\, \mathcal{P}_{i}$ and $p_d\, \mathcal{P}_{j}$. Equation \eqref{EQ:Derivative=0} shows that the mean photon pair number $\mu$ that maximises the visibility \eqref{EQ:Vmain} for the pair of detectors $(i,j)$ is such that the product of the individual detection probabilities in these detectors equals the probability to have a coincidence count between noise events on the same detectors. It is worth remembering that this conclusion is valid for low mean number of pairs $\mu$ when compared to the number of photon pair modes. Working with low $\mu$, we can approximate the average number of detected photons on detectors $i$ and $j$  as $p_d \mu  \mathcal{P}_{i}$ and $p_d \mu \mathcal{P}_{j}$, respectively, whereas the average number of detected photons on each detector array is $p_d \mu$. Using the condition for maximum visibility \eqref{EQ:Derivative=0}, the average number of detected photons in the full detector array is
\begin{equation}\label{EQ:Derivative=02}
p_d \mu \approx \frac{p_n}{\sqrt{\mathcal{P}_{i} \mathcal{P}_{j}}}.
\end{equation}

For most of the correlated states, the marginals $\mathcal{P}_i$ and $\mathcal{P}_j$ evaluated at the peaks of the joint detection probability are equivalent. Considering $\mathcal{P}_i=\mathcal{P}_j$, equation \eqref{EQ:Derivative=0} shows that  the maximum visibility is achieved by adjusting $\mu$ such that the mean number of detected photons (in $i$ or $j$) equals the mean number of noise events $p_n$ on the detectors. Since the noise probability on most of the available detector arrays is of the order of $p_n \lesssim10^{-2}$, we are considering an average number of detected photons of the same order: $p_d \mu \mathcal{P}_i=p_d \mu \mathcal{P}_j =p_n \lesssim10^{-2}$. This conclusion provides a guide to obtain the maximum visibility in measurements of intensity correlations of photon pairs. After characterisation of the noise probability $p_n$ of the detector array, the photon pair flux (and/or the exposure time) has to be adjusted such that the mean number of photo-detections on the detectors for which the correlation function is peaked is equivalent to $p_n$. 
Strictly speaking, as the mean number of photons on the detectors increases, the probability of having a noise event decreases, according to equation \eqref{EQ:AveNumEvents}. Nevertheless, since the noise probabilities on the available detector arrays are low, typical experiments can be operated with low mean number of photons. Then, as long as the approximation of the mean number of photo-detections as $p_d\mu \mathcal{P}_i$ is valid, this conclusion holds. It is also interesting to see that, when working with the condition for maximum visibility \eqref{EQ:Derivative=0}, each term in the reference correlation function $\bar{G}^{(2)}_{i j}$ becomes equivalent to $p_n^2$, which is the contribution of the coincidence counts between noise events. Using the mean number of photo-detections as $p_d\mu=p_n/\mathcal{P}_{i}=p_n/\mathcal{P}_{i}$, we obtain $\bar{G}^{(2)}_{i j}=4p_n^2$.

\subsection{Example: multimode correlations with a uniform marginal detection probability distribution }

\begin{figure}[t]
\begin{center}
\includegraphics[width=8.5cm]{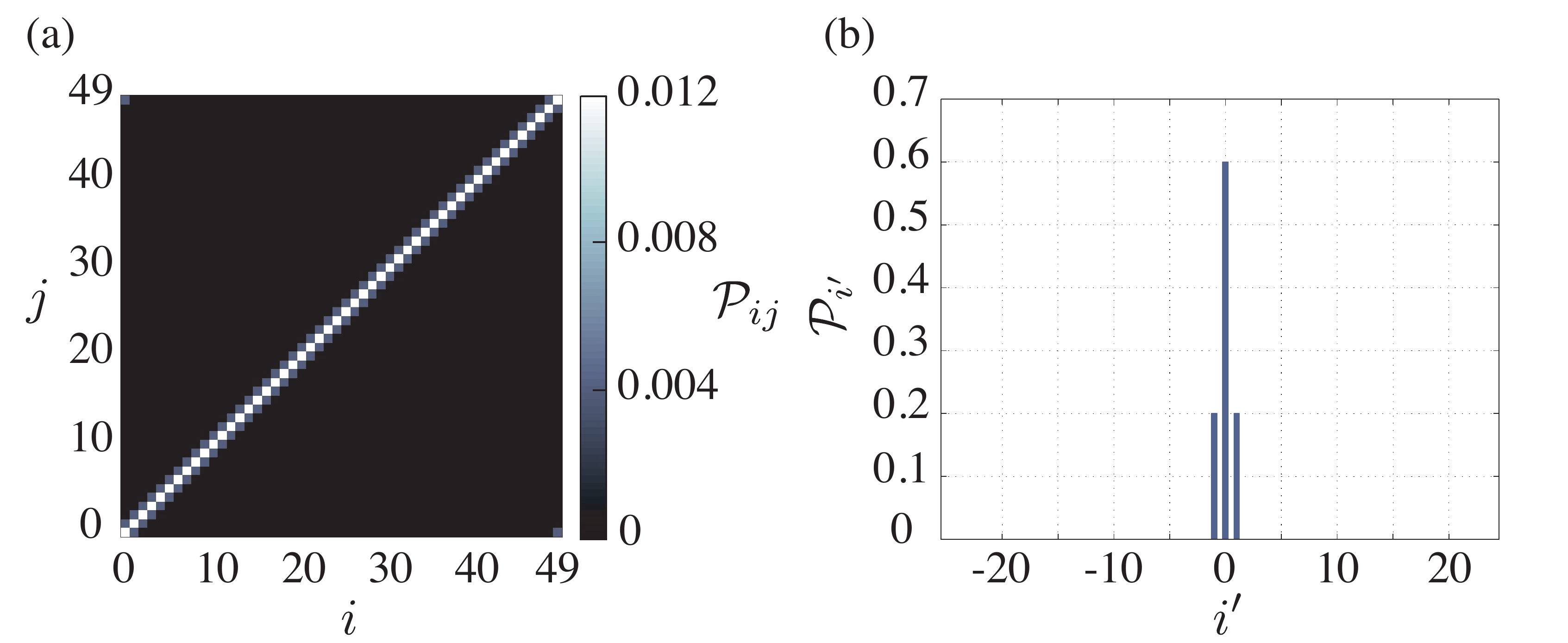} 
\caption{(Color online) (a) Joint $\mathcal{P}_{ij}$ and (b) marginal $\mathcal{P}_{i'}$ probability distributions for the state \eqref{EQ:StateEx1} with $c=0.6$ and $D=50$. Here the $i' \equiv i-j$ is the difference of the detector labels.}
\label{Fig:FlatState}
\end{center}
\end{figure}

\begin{figure}[t]
\begin{center}
\includegraphics[width=8.4cm]{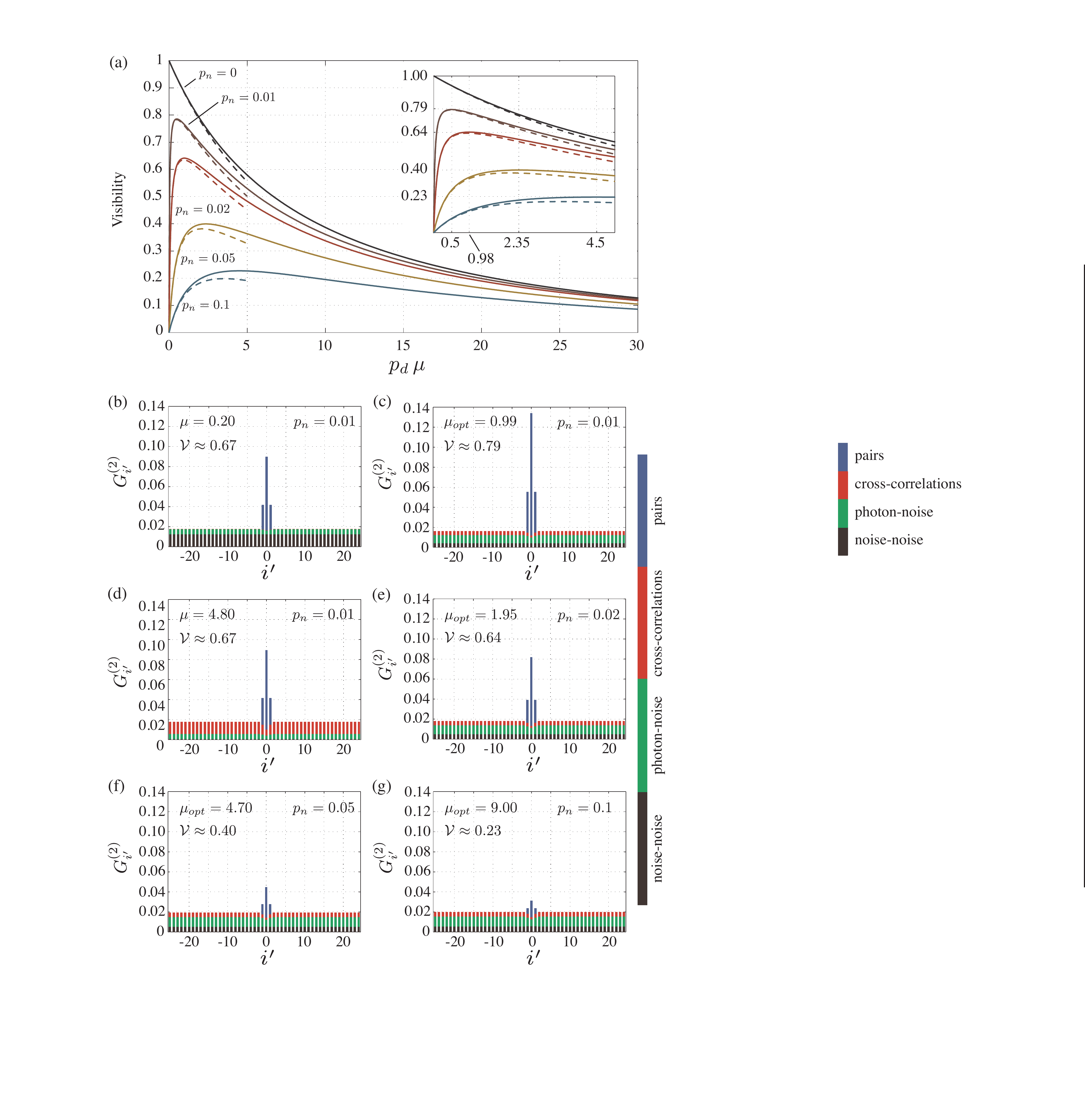} 
\caption{(Color online) (a) Visibility of the marginal correlation function of the state \eqref{EQ:StateEx1} for $c=0.6$, $D=50$ and $p_d=0.5$ for many values of the noise probability $p_n$. The solid lines are plots of the visibilities as calculated from equation \eqref{EQ:G2ALL2} whereas the dashed lines correspond to the quadratic approximation given in equation \eqref{EQ:Vmain}. 
The stacked bar graphs displayed from (b) to (g) show the four different contributions to the measured correlation function in different colors, as indicated in the legend.
The correlation functions calculated using the condition for maximum visibility are shown in (c) for $p_n=0.01$, (e) for $p_n=0.02$, (f) for $p_n=0.05$ and (g) for $p_n=0.1$. A comparison of the correlation function for $p_n=0.01$ with different mean number of pairs is shown in (b), (c) and (d). }
\label{Fig:VisFlatState}
\end{center}
\end{figure}

As a simple example, let us consider the two-photon entangled state 
\begin{equation}\label{EQ:StateEx1}
|\Psi \rangle_{12} = \sum_{i=0}^{D-1} \, \sum_{j=0}^{D-1} \, \sqrt{\frac{c_{ij}}{D}} \, | i \rangle_1 | j \rangle_2,
\end{equation}
with the coefficients $c_{ij}$  given by
\begin{equation}\label{EQ:CoeffEx1}
c_{ij}= c \, \delta_{ij} +  \frac{1-c}{2} \, \delta_{i (j+1)} +  \frac{1-c}{2} \, \delta_{(i+1)j},
\end{equation}
where the sums in the indices of the Kronecker deltas are modulo D; $\delta_{0D}=\delta_{D0}=1$. Here $|i\rangle_1$  $(|j\rangle_2)$ represent a single-photon state in the detector mode $i$ ($j$) of detector array $1$ ($2$), and we assume each detector array is equipped with $D$ detectors. For $c=1$, the state \eqref{EQ:StateEx1} is the maximally entangled state in dimension $D$. The two-photon state \eqref{EQ:StateEx1} generates a joint detection probability distribution $\mathcal{P}_{ij}=c_{ij}/D$ over the detector arrays. Figure \ref{Fig:FlatState}-a shows the joint probability distribution $\mathcal{P}_{ij}$ for the state \eqref{EQ:StateEx1} with $c=0.6$ in $D=50$ dimensions. The marginal probability distribution $\mathcal{P}_{i'}$ as a function of the difference of the detector labels is plotted in figure \ref{Fig:FlatState}-b, where we define $i'\equiv i-j$. The intensity correlations of this two-photon state are such that the photons are detected with $60\%$ probability on detectors with same label ($i'=0$) and with $40\%$ probability on detectors whose labels differ by one ($i'=\pm1$).

The correlation function $G^{(2)}_{ij}$ for this state is particularly easy to analyse, since the marginal probability distributions $\mathcal{P}_i=\mathcal{P}_j=1/D$ are constant, {\it i.e.} independent of $i$ or $j$. This means that the mean number of photons in each detector is the same, given by $\mu/D$. Using the condition for maximum visibility given in equation \eqref{EQ:Derivative=02}, we get a mean number of photo-detections on each detector array given by $p_d\mu=p_nD$, which is equivalent to the average number of noise events in the full detector array. In this case, the optimum mean number of photon pairs can be adjusted by matching the mean number of photo-detections and noise events in the full detector array. 

We used equation \eqref{EQ:G2ALL2}  and its quadratic approximation given by equation \eqref{EQ:G2AllApprox} to calculate the intensity correlation function of the state \eqref{EQ:StateEx1}. Figure \ref{Fig:VisFlatState}-a shows the visibility of  $G^{(2)}_{i'}$ as a function of $p_d\mu$ for many values of the noise probability $p_n$ and with a detection probability of $p_d=0.5$. The intensity correlation function $G^{(2)}_{i'}$, which is associated with the measurement of the non-local marginal probability distribution $\mathcal{P}_{i'}$, is obtained from $G^{(2)}_{ij}$ by summing over the diagonal defined by the variable $j'\equiv i+j$. The solid lines are plots of the visibilities as calculated from equation \eqref{EQ:G2ALL2} and the dashed lines correspond to the approximation given in equation \eqref{EQ:Vmain}. We see that as the noise probability increases the mean number of photon pairs to obtain the maximum visibility increases, whereas the maximum achievable visibility gets smaller. The perfect visibility $\mathcal{V}=1$ is only achievable with an ideal detector ($p_n=0$) and in the limit of very small number of photons $\mu \rightarrow 0$. For $p_n=0.01$, the optimum visibility of $\mathcal{V}\approx 0.79$ is achieved with $\mu_{opt}\approx 0.99$, which is in agreement with the prediction of equation \eqref{EQ:Derivative=02} of $\mu_{opt}= Dp_n/p_d =1$. As the noise probability gets larger, the prediction of $\mu_{opt}$ from equation \eqref{EQ:Derivative=02} deviates from the exact solution, as it can be seen from the inset of figure \ref{Fig:VisFlatState}-a. For $p_n=0.1$ we have $\mu_{opt}\approx 9$, whereas equation \eqref{EQ:Derivative=02} predicts $\mu_{opt}=10$.

In figures \ref{Fig:VisFlatState}-b to \ref{Fig:VisFlatState}-d we show the correlation function for $p_n=0.01$ calculated for three different values of $\mu$. Since the state \eqref{EQ:StateEx1} generates a uniform photon flux on all detectors, the background of the correlation function is also uniform. Experimentally this is a convenient situation as the peak of the correlation function becomes easily identifiable from the background, and the visibility can be experimentally defined without ambiguity. In this case, calculating the visibility according to our definition \eqref{EQ:Vdef} is equivalent to taking the peak and the background values as the {\it maximum} and {\it minimum} intensities, respectively. Using the optimum mean number of photon pairs $\mu_{opt}$ (figure \ref{Fig:VisFlatState}-c), the background of the correlation function has equal contributions of noise-noise correlations and cross-correlations, whereas the photon-noise contribution is twice that of the noise-noise contribution. In figures \ref{Fig:VisFlatState}-b and  \ref{Fig:VisFlatState}-d the mean number of photon pairs $\mu$ were chosen to give the same visibility, but the contributions of the different coincidence count terms are different in each case. Using $\mu< \mu_{opt}$ (figure \ref{Fig:VisFlatState}-a) the dominant  contribution comes from the noise-noise correlations, whilst the cross-correlations become dominant for $\mu > \mu_{opt}$ (figure \ref{Fig:VisFlatState}-d). In figures \ref{Fig:VisFlatState}-c and \ref{Fig:VisFlatState}-e to \ref{Fig:VisFlatState}-g we compare the optimal correlation functions for a range of $p_n$ values, finding a reduced maximum visibility for increasing values of the noise probability. For example, by increasing the noise probability by a factor of 10, from $p_n=0.01$ to $p_n=0.1$,  the maximum achievable visibility decreases from $79\%$ to $23\%$.

\section{Experiment: far-field intensity correlations in SPDC}
\label{sec:Experiment}

\begin{figure}[t]
\begin{center}
\includegraphics[width=9cm]{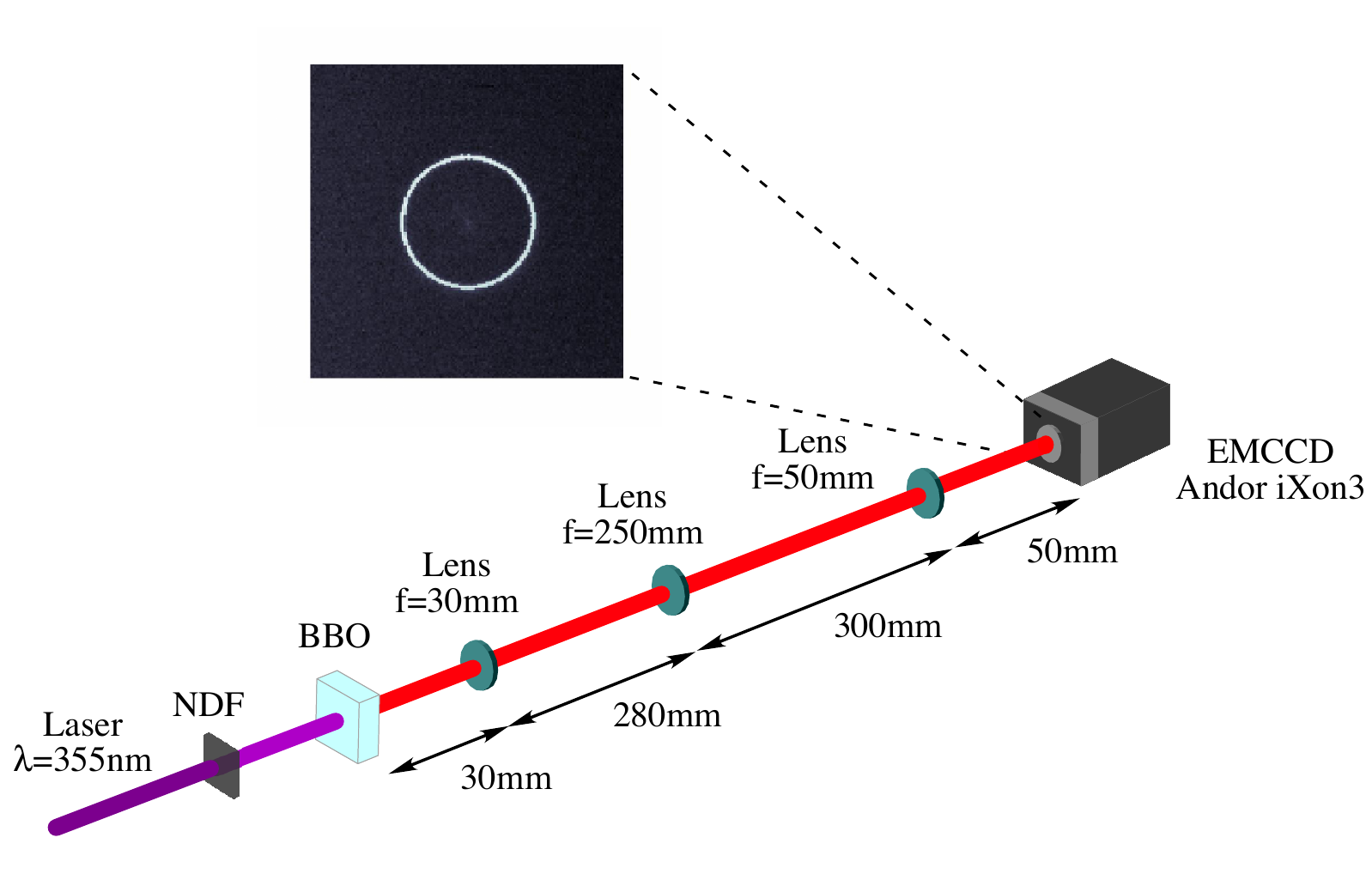} 
\caption{(Color online) Experimental setup used to measure far-field intensity correlations with an EMCCD camera. The inset shows the far-field intensity distribution of the down-converted fields as seen on the camera. NDF is a neutral density filter.}
\label{Fig:Exp Setup}
\end{center}
\end{figure}

We have experimentally tested the conditions described in section \ref{sec:IC} by measuring far-field intensity correlations of the photon pairs from SPDC with an EMCCD camera. A diagram of our experimental setup is shown in figure \ref{Fig:Exp Setup}. A $150$mW high repetition rate laser $@ 355$nm is attenuated with a neutral density filter and used to pump a $3$mm long BBO crystal cut for type-I phase-matching. The down-converted fields were detected through a $10$nm wide (FWHM) interference filter centered at $710$nm placed in front of the camera. A Fourier lens system with effective focal length of $f_e=6$mm was used to produce the far-field intensity distribution of the down-converted fields on the EMCCD, as shown in figure \ref{Fig:Exp Setup}.

Our experimental setup for measuring intensity correlations does not exactly correspond to the configuration described in section \ref{sec:IC}, as we detect both photons on the same detector array. In the configuration analysed previously, ``$i$"  and ``$j$" were used to label detectors in different detector arrays, each of which used to detect one of the photons from a pair. Detecting each photon in a different detector array enables the labeling of the photons such that they are distinguishable: a photon occupying the detector mode $i$ can be called {\it signal} and a photon occupying the detector mode $j$ can be called {\it idler}. On the other hand, detecting signal and idler photons on the same detector array makes them indistinguishable: the occupation probability of the detector modes $i$ ($j$) have contribution of both signal and idler photons. As a consequence, it is impossible to measure the full set of correlations between the photon modes. In the configuration shown in figure \ref{Fig:Exp Setup}, the coincidence counts between detectors $i$ and $j$ are associated with the joint detection probability $\mathcal{P}_{ij}$ and $\mathcal{P}_{ji}$, and the result is that the measured correlation function assumes the mirrored form given by $G^{(2)}_{ij}=G^{(2)}_{ji}$. 

Another consequence of detecting both photons on the same detector array is that, unlike the situation described earlier, we also measure signal-signal and idler-idler correlations, increasing the number of cross-correlations. The scaling of the four coincidence counts terms in equation \eqref{EQ:CC4terms} with the mean number of pairs $\mu$ changes but, interestingly, the scaling with the average number of {\it detected} photons is preserved. With the two detector arrays configuration of section \ref{sec:IC}, the average number of detected photons on pixels $i$ and $j$ are $p_d\mu \mathcal{P}_i$ and $p_d\mu \mathcal{P}_j$, respectively. By using only one detector array with the same mean number of pairs $\mu$, these numbers double. Here we have assumed that signal and idler photons are completely indistinguishable, such that their marginal detection probabilities are the same. By adapting the occupation probabilities given in equations \eqref{EQ:ModeNumber}, \eqref{EQ:mucrossij}, \eqref{EQ:muYesBar}, \eqref{EQ:muBarYes} and \eqref{EQ:muBarBar}, it is straightforward to show that the condition for maximum visibility using only one detector array is $(2p_d\mu \mathcal{P}_i)(2p_d\mu \mathcal{P}_j)\approx p_n^2$. Comparing this condition for maximum visibility with equation \eqref{EQ:Derivative=0}, we see that the optimum mean number of pairs when working with only one detector array is half of that required in the case where two detector arrays are employed. Nevertheless, in both situations, the maximum visibility is achieved by matching the photo-detection probabilities with the noise probabilities.

The intensity distribution pattern shown in figure \ref{Fig:Exp Setup} is the result of an accumulation of many down-conversion emissions. This ring-shaped structure represents the full set of modes in which each single down-converted photon can be detected on the EMCCD. In order to measure the intensity correlations of the photon pairs, {\it i.e.} the probability distribution for joint detections of the photons, we need a set of spatially sparse frames containing only a few photo-detections.  The mean number of photon pairs per frame was adjusted by controlling the exposure time $\tau_e$ of the EMCCD and adjusting the attenuation of the pump laser by way of neutral density filters (NDF) with different optical densities (OD). We used four different combinations of NDF, giving optical densities of $1.3$, $1.4$, $1.6$ and $1.8$. For each combination of NDF, the intensity correlations were measured using many exposure times. The range of exposure times used was from $1$ms to $50$ms for the two weakest pump beams (OD$=1.6$ and OD$=1.8$) and from $0.1$ms to $50$ms for the two strongest pump beams (OD$=1.3$ and OD$=1.4$).

The EMCCD camera used in our experiment was an Andor iXon3 with a back-illuminated sensor containing an array of $512 \times 512$ pixels of size $s_p=16  \mu$m. In our measurements, the EMCCD sensor was air-cooled to $-85^\circ$C and the gain was set to maximum. Other operational parameters used were horizontal pixel shift read-out rate of $10$MHz and a vertical pixel shift every $0.3\mu$s.  We selected an annular region of interest (ROI) on the camera comprising $660$ pixels on the CCD chip. This annular ROI was $4$ pixels wide, with inner and outer radii given respectively by $24$ and $28$ pixels, and was chosen to enclose the down-conversion ring on the camera. 
With the operational parameters of the EMCCD set, we characterised the level of noise in the annular ROI  for every exposure time used. The noise characterisation consisted of a statistical analysis of the pixel outputs with the camera settings of interest but with the shutter of the camera closed \cite{Lantz08,Jedrkiewicz12}. By taking a series of frames {\it in the dark}, we calculated the average signal output for every pixel that was then used as background subtraction in each acquisition. The statistical distribution of these background subtracted pixel outputs taken in the dark was used to choose a threshold for single-photon discrimination. Based on the chosen threshold, we calculated the probability $p_n$ that a signal output would exceed the threshold in the absence of light. As the main source of noise in our measurement is due to clock induced charges during the read-out of the CCD chip, 
we found that the mean number of noise events per frame does not strongly depend on the exposure time used. For a range of exposure times varying from $0.1$ms to $10$ms, the average number of noise events measured in the annular ROI was $6.2$, giving a noise probability of $p_n = 6.2/660 \approx 0.94 \times 10^{-2}$. Increasing the exposure time beyond $10$ms slightly increased the average number of noise events: for $\tau_e=50$ms we found $p_n=1.47 \times 10^{-2}$.

\begin{figure}[t]
\begin{center}
\includegraphics[width=8.5cm]{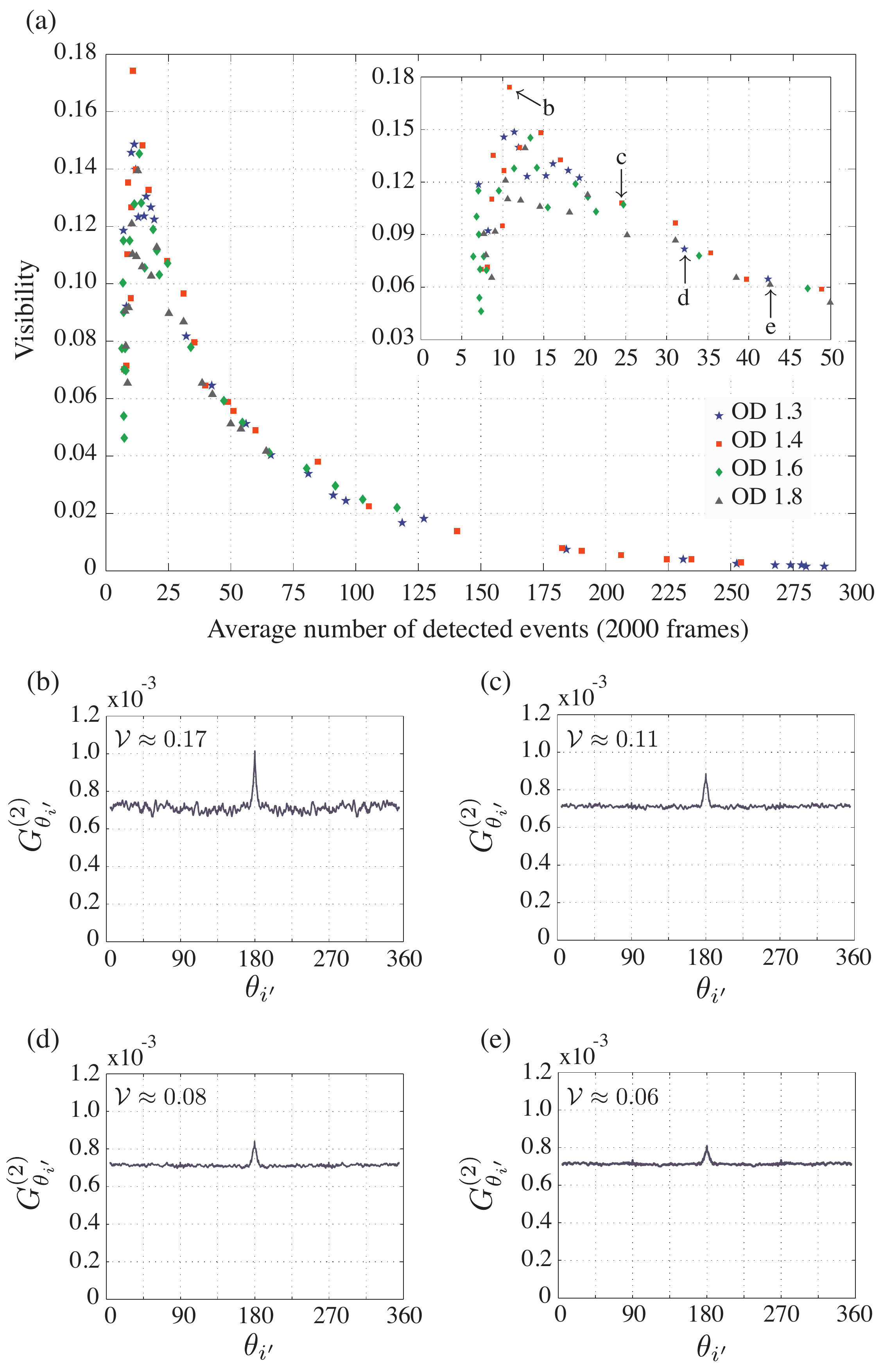} 
\caption{(Color online) (a) Visibility of the measured correlation functions as a function of the average number of detected events. Samples of the measured correlation function for the points indicated in the inset are displayed   on (b), (c), (d) and (e). }
\label{Fig:ExpResults}
\end{center}
\end{figure}

The transverse correlations of the photon pairs from SPDC arise from energy and momentum conservation of the fields involved in the down-conversion process \cite{Saldanha2013}. In the far-field, the transverse positions of the photons are anti-correlated \cite{Howell04}. The strength of the far-field intensity correlations is determined by the width of the Gaussian pump laser and on the optical Fourier system used to produce the far-field distribution on the camera. For our system, we calculate the width of the intensity correlations to be $f_e/(k\sigma_p) \approx 1\mu$m, where $k$ is the wave-number of the down-converted fields and $\sigma_p$ is the standard deviation of the Gaussian pump beam. As the correlation length of the photon pairs is much smaller than the pixel size $s_p=16\mu$m, the photo-detections corresponding to photons of the same pair happen mostly on pixels that are located diametrically opposed in the annular ROI. 

For each exposure time and pump beam power, we took a series of $2000$ frames and counted the coincidences as a function of the pixel labels. Each pixel within the annular ROI was labeled according to its polar coordinates ($r_i,\theta_i$), where the origin of the reference system was chosen to be the center of the ring-shaped far-field intensity distribution. We then calculated the intensity correlation function as a function of the difference of the pixel angles ($\theta_{i'}\equiv\theta_i- \theta_j$). Since the photon pairs are anti-correlated, the measured correlation function is peaked at $\theta_{i'}=180^{\circ}$ \cite{Oemrawsingh02}. In order to reduce  pixelation effects in our measurements the angles were binned in steps of $2^{\circ}$. The visibilities calculated from  the measured correlation functions are shown in figure \ref{Fig:ExpResults}-a as a function of the average number of {\it detected} events (per frame) in the full annular ROI. In figures \ref{Fig:ExpResults}-b to \ref{Fig:ExpResults}-e we show four samples of the measured correlation functions for the points indicated in the inset of figure \ref{Fig:ExpResults}-a. It is important to note that the average number of detected events contain contributions from the photo-detections and noise events. For low mean number of pairs $\mu$, the average number of detected events can be written as $6.2+2p_d\mu$.
As the photon flux on the detector array decreases, the average number of measured events approaches the number of events expected from noise ($\approx 6$ events), as can be seen in figure \ref{Fig:ExpResults}-a. 

For different pump powers, the same number of detected events is achieved by using different exposure times. As the noise probability is approximately the same for every exposure time used, the visibility curve for the four sets of measurements (each of the four different pump powers) displays the same behaviour when plotted against the average number of events. The maximum visibility was achieved for a number of detected events in a range from 10 to 15, which is consistent with the condition for the maximum visibility stated above. Our experimental setup is similar to the example discussed in section  \ref{sec:IC}, since the marginal probability distributions for each single-photon are constant, {\it i.e.} independent of the angle $\theta$. We divided the annular detector array into $180$ bins $2^{\circ}$ wide, but due to the symmetry of the detection system the number of pixels within each of the $180$ bins is not the same, varying mostly between $3$ and $4$. On average each bin contains $660/180\approx 3.7$ pixels, and we write the marginal probability distribution of the photons as $\mathcal{P}_{\theta_i}=\mathcal{P}_{\theta_j} \approx 1/180$. The average probability of detecting a noise event in each bin is then given by $3.7p_n$. Calculating the average number of detected photons in the annular ROI that is associated with the condition for maximum visibility we get $2p_d\mu \approx 3.7p_n/\sqrt{\mathcal{P}_{\theta_i} \mathcal{P}_{\theta_j}}=660p_n$, which is equivalent to the average number of noise events in the detector array. As we mentioned previously, this conclusion is valid for low noise probabilities $p_n$, so that the mean number of photo-detections required to match this condition is also low.

Another interesting feature that we observe in our measurements is the difference in the statistical fluctuation of the many intensity correlation functions measured. The statistical sampling of each measured point of the correlation function depends on the average number of coincidence counts measured for that point. As the number of frames used in the calculation of each correlation function was the same ($2000$), the statistical sampling is higher for the correlation functions measured with higher average number of pairs. For example, the statistical fluctuations of the correlation function displayed in figures \ref{Fig:ExpResults}-b are higher than that of figure \ref{Fig:ExpResults}-e. Although the correlation function in figure \ref{Fig:ExpResults}-b has a higher visibility, since it was measured with a mean number of pairs close to the condition for maximum visibility, the curve in figure \ref{Fig:ExpResults}-e displays smaller fluctuations. The average number of detected events associated with figures  \ref{Fig:ExpResults}-b and \ref{Fig:ExpResults}-e are $10.8$ and $42.6$, respectively. So in order to obtain statistical fluctuations comparable with that of figure \ref{Fig:ExpResults}-e, approximately $8000$ frames would have to be taken with the settings of figure \ref{Fig:ExpResults}-b.

\section{Conclusions}
\label{sec:Conclusions}

Intensity correlation measurements are ubiquitous in SPDC experiments. With the advance of the modern detector arrays, especially single-photon sensitive cameras, multimode coincidence detection of the photon pairs is becoming a more attractive alternative than the traditional scanning detection systems. Modern single-photon sensitive CCD cameras available off-the-shelf can provide an array of up to one million individual detectors that are sensitive to single-photons with high quantum efficiencies. Whilst the noise level in these cameras is still higher than in the traditional single-photon avalanche diodes, it is already low enough to enable coincidence detection in SPDC  experiments, as it has been demonstrated in a number of recent works~\cite{Edgar12, Moreau12,Aspden12,Fickler12}. 

Here we provided a detailed analysis of the use of detector arrays for multimode detection of intensity correlations in the single-photon regime. We considered the coincidence count distributions generated by the photon pairs from SPDC in a multimode configuration, providing ready-to-use expressions for the measured correlation function in terms of the detector parameters and on the mean number of detected modes. The correlations introduced by the simultaneous detection of multiple photon modes and noise events were studied and we showed how each of these contributions scales with the number of detected modes. For noise levels smaller than one noise event per every 100 pixels, the conditions for the maximisation of the visibility of the measured correlation function is achieved by matching the photo-detection rate with the noise event rate on the detectors for which the correlation function is peaked. This condition can be translated into the average number of detected modes in the full detector array, and provides a guide for the optimisation of coincidence counts measurements with detector arrays.

\acknowledgments
M.J.P. thanks the Royal Society, the Wolfson Foundation and the US DARPA/DSO InPho program. D.S.T., M.P.E., F.I. and G.S.B. acknowledge the financial support from the UK EPSRC. The authors would like to thank R. S. Aspden for the careful reading of this manuscript. D.S.T. acknowledges A. Z. Khoury for useful discussions.



\end{document}